\documentclass[fleqn,usenatbib]{mnras}
\usepackage[dvipsnames]{xcolor}
\usepackage[english]{babel}
\usepackage[autostyle, english = british]{csquotes}
\usepackage{amsmath}
\MakeOuterQuote{"}
\usepackage{hyperref}
\hypersetup{
    colorlinks=true,
    linkcolor=ForestGreen,
    filecolor=ForestGreen,      
    urlcolor=ForestGreen,
    citecolor=ForestGreen,
    }

\usepackage{newtxtext,newtxmath}
\usepackage[T1]{fontenc}

\DeclareRobustCommand{\VAN}[3]{#2}
\let\VANthebibliography\thebibliography
\def\thebibliography{\DeclareRobustCommand{\VAN}[3]{##3}\VANthebibliography}

\usepackage{graphicx}
\usepackage{amsmath}	
\definecolor{scarlet}{HTML}{bb0000}

\newcommand{\hst}{{\it HST}}
\newcommand{\gaia}{{\it Gaia}}

\newcommand{\hop}{{\sc h1p}}

\definecolor{scarlet}{RGB}{187, 0, 0}

\title[HubPUG]{\color{scarlet}HubPUG: \textcolor{gray}{Proper Motions} for \textcolor{gray}{Local Group Dwarfs} observed with \textcolor{gray}{HST} utilizing \textcolor{gray}{Gaia} as a \textcolor{gray}{Reference Frame}}

\author[J. T. Warfield et al.]{Jack T. Warfield,$^{1}$\thanks{E-mail: jtw5zc@virginia.edu}
Nitya Kallivayalil,$^{1}$
Paul Zivick,$^{2,3}$
Tobias Fritz,$^{1}$
Hannah Richstein,$^{1}$\newauthor
Sangmo Tony Sohn,$^{4}$
Andr\'es del Pino,$^{5,4}$
Alessandro Savino,$^{6}$
and Daniel R. Weisz$^{6}$
\\
$^{1}$Department of Astronomy, The University of Virginia, 530 McCormick Road, Charlottesville, VA, 22904, USA\\
$^{2}$Mitchell Institute for Fundamental Physics and Astronomy,
Department of Physics and Astronomy,
Texas A\&M University,
College Station, TX, 77843, USA\\
$^{3}$Department of Physics and Astronomy,
Texas A\&M University,
College Station, TX, 77843, USA\\
$^{4}$Space Telescope Science Institute,
3700 San Martin Drive,
Baltimore, MD, 21218, USA\\
$^{5}$Centro de Estudios de F\'isica del Cosmos de Arag\'on (CEFCA), Unidad Asociada al CSIC, Plaza San Juan 1, E-44001, Teruel, Spain\\
$^{6}$Department of Astronomy,
University of California,
Berkeley, CA, 94720, USA
}

\date{Accepted XXX. Received YYY; in original form ZZZ}

\pubyear{2022}

\begin{document}
\label{firstpage}
\pagerange{\pageref{firstpage}--\pageref{lastpage}}
\maketitle

\begin{abstract}
We present the method behind {\sc HubPUG}, a software tool built for recovering systemic proper motions (PMs) of {\it Hubble Space Telescope} (\hst) fields with two epochs of observations by utilizing stars observed by \gaia\ as a foreground frame of reference.
\hst\ PM experiments have typically relied on the use of distant background galaxies or quasi-stellar objects (QSOs) as stationary sources against which to measure PMs. Without consistent profiles, background galaxies are more difficult to centroid, but benefit on-aggregate from their large numbers. QSOs, though they can be fit with stellar point-spread functions, are sparse, with most fields containing none.
Historically, the use of stars as references against which to measure PMs would have been difficult because they have individual PMs of their own.
However, \gaia\ has now provided positions and PMs for over 1.4 billion stars, which are much more likely to be well-imaged in the fields around targets versus background sources, have predictable stellar profiles, and require less observing time per-image for good signal-to-noise.
This technique allows us to utilize the power of \gaia\ to measure the PM of targets too faint for \gaia\ to observe itself.
We have recovered PMs for the Milky Way satellites Sculptor and Draco with comparable uncertainties over \hst-only and \gaia-only measurements, limited primarily by the current capabilities of the \gaia\ data. We also show the promise of this method for satellites of M31 with a new PM measurement for Andromeda VII.
\end{abstract}




\section{Introduction} \label{sec:intro}

On the scale of the Local Group (LG) -- the Andromeda (M31) and Milky Way (MW) systems -- the motions that we measure of galaxies on the sky (their proper motions; PMs) can tell us a great deal about problems in near-field $\Lambda$CDM cosmology and about galaxy evolution.
For instance, PMs have allowed us to probe the apparent existence of satellite planes \citep[e.g.,][]{LB1976, Kunkel1976, Kroupa2005}.
Both the MW's Vast Polar Structure \citep[e.g.,][]{Pawlowski2012, Sohn+2017, Fritz+2018, Pace2019, Pawlowski2020} and the Great Plane of Andromeda \citep{sohn+2020, Pawlowski2021} have been explored in recent years using data from \hst\ and the \gaia\ Survey \citep{gaia1}.
Additionally, the nature of dark matter (DM) and halo substructure is also expected to be clarified with more precise PM measurements. For example, PMs could help break the degeneracy between the density profile and stellar anisotropy, allowing for better characterization of the central DM density \citep{Strigari2007a, Read2021, Guerra2022}.

As pertains to galaxy evolution, with \gaia-based PMs we have learned more about accretion and merger events in the assembly of the MW \citep[e.g.,][]{Belokurov2018, Haywood2018, Helmi2018, Venn2018} and the Magellanic System  \citep[e.g.,][]{Kallivayalil2018, Fritz+2018, Simon2019, Erkal2020, Patel2020}. Pairing PM measurements with \hst\ star-formation histories (SFHs) has established the relationship between dwarf galaxies and reionization \citep[e.g.,][]{brown2014, weisz2014a, boylankolchin2015} and allowed us to explore environmental influences on star formation in the faintest of galaxies \citep{Sacchi2021}.

While more than 75\% of the observed MW satellites are being studied in detail with current PM information \citep[e.g.,][]{McConnachie2020,li+2021,Pace2022}, to date, only two M31 dwarf satellites (NGC 147 and NGC 185, by \citealt{sohn+2020}) have \hst-based PM measurements.
This can partially be attributed to how time-expensive these experiments have historically been. 
To measure the PM of distant objects such as dwarf galaxies, at least two epochs of images need to be taken with a time-baseline $t\gtrsim5$ years, and these images must have the precision to be able to resolve motions $\lesssim(0.01{\rm\,mas/yr}) \times t$.
The use of \hst's Advanced Camera For Surveys/Wide Field Channel (ACS/WFC; and/or \hst's Wide-Field Camera 3 [WFC3]) is therefore crucial for its ability to capture high-precision astrometry, which can be further enhanced through observational strategies, e.g., dithering to fully sample a star's point-spread function (PSF).

In order to measure the PM of a field of stars, a common frame-of-reference is needed across observational epochs. Traditionally, background galaxies and/or quasi-stellar objects (QSOs, or "quasars") -- objects distant enough to be approximately stationary between epochs -- are used as references against which it is possible to track the motion of the stars in a target galaxy.
Therefore, exposure times for observations not only need to be planned to obtain adequate signal-to-noise (S/N) for the dwarf galaxy's stellar component, but also for these background sources. In the case of \cite{sohn+2020}, each of the first- and second-epoch images used for their targets had exposure times of $\sim$1400 seconds in the ${\rm F606W}$ filter (with 20 images being taken for the first epoch and 8 for the second). The analysis and astrometric reduction for these experiments can also be intensive, requiring careful treatment of instrumental geometric distortions, cross-detector PSFs, and centroiding of background sources.

However, the data from the \gaia\ survey, including the recent Early Data Release 3 \citep[eDR3;][]{gaiaedr3} and Data Release 3 \citep[DR3;][]{gaiadr3}, have offered a new opportunity to greatly expand the sample of dwarf galaxies with resolved PMs. \gaia\ DR3 provides sky positions and PMs for over 1.4 billion stars. Though a majority of these sources are Galactic, the catalog does include data for bright stars in the MW halo and its satellites. Therefore, a target galaxy's systemic PM can be measured by averaging the stellar PMs measured by \gaia\ for stars that are likely members of the target galaxy (we abbreviate this method as "\gaia-only"). This has been done by, e.g., \cite{Fritz+2018} with \gaia\ Data Release 2 \citep{gaiadr2} and by \cite{Battaglia+2021}, \cite{li+2021}, and \cite{MartinezGarcia+2021} with eDR3.
The value of the \gaia\ data and this method is hard to overstate: \cite{Battaglia+2021} provides PM measurements for 74 LG satellites, and \cite{li+2021} recovers full orbital parameters for 46 MW dwarfs. 

Despite its unquestionable utility for directly recovering systemic PMs, the \gaia-only method is only possible for galaxies with an adequate number of members bright enough such that they have stellar PMs measured by \gaia.
Therefore, this method is limited to targets in the MW's halo and the brightest of M31's satellites (e.g., M33).
However, as well as being powerful in its own right, the \gaia\ data may also serve as a component in making high-precision measurements of satellite PMs when combined with data from \hst.

Historically, stationary background sources have been used, in part, because of the difficulty of using foreground sources that have PMs of their own. However, this comes with the trade-off that the profiles of background galaxies are not uniform (as opposed to the consistent stellar PSF), and so the positions of these sources are more difficult to measure.
QSOs, also being point-sources, can be fit using stellar profiles, but these sources are very sparse in the backgrounds of target fields (most fields will have none). On the other hand, while foreground MW stars have motions of their own, these motions have now been measured and reported using \gaia. Thus, notwithstanding whether or not a star is a member of a target galaxy, these individual PMs can be theoretically subtracted away after comparing how these foreground stars move in reference to a target galaxy over time, revealing a measurement of the target's systemic PM. 

Because these foreground stars will be, on average, brighter than the target's member stars, using foreground sources may allow for less time-expensive experiments, as we will not need to consider added exposure time to increase the S/N of faint background galaxies. In addition, because foreground stars can be fit with the same stellar PSFs as the stars of target galaxies, we should be able to obtain more precise positions of these references in each of the \hst\ epochs versus background galaxies. This may especially be relevant for combining observations from different filters and/or instruments, where the the cross-filter centroiding of stellar profiles is much more robust than the profiles of extended objects. Therefore, we would expect this method to be probing the limits of the precision possible with \hst\ astrometry. In fact, though the most significant limits should be from the uncertainty associated with \gaia's stellar PM measurements and from the number of foreground stars in an \hst\ field, the addition of \hst\ data has the potential to provide more precise estimates for the PMs of targets that already have purely \gaia-derived measurements.

There are already a few examples in the literature of attempts at combining \hst\ and \gaia\ data for PMs. \cite{casdin+2021hubgaia} combined observations from multiple \hst\ Wide Field and Planetary Camera 2 epochs and \gaia\ eDR3 to obtain a PM of Leo I, using the average "offset" between the \gaia-reported stellar PMs and the PMs of these stars as implied by their movements between the \hst\ epochs and the \gaia\ frame-of-reference as a correction to the \hst-derived motions. Their team recovered values within $1\sigma$ of the multi-study average value calculated from the literature. Similarly, the {\sc GaiaHub} tool \citep{gaiahub} allows \gaia\ to be used as a second epoch to an \hst\ observation to find relative PMs and velocity dispersions within fields. This is similar to previous work that was done recovering PMs for Sculptor and Draco by \cite{Massari+18} and \cite{Massari+20}, respectively, where the stellar positions from \gaia\ DR2 were used as a second epoch against the first epoch \hst-derived positions of these same stars to estimate absolute PMs.

In this paper, we accelerate this exciting effort to unify data from \hst\ and \gaia, presenting the method behind our software package {\sc HubPUG} (Hubble Proper motions Utilizing Gaia), an application capable of recovering the systematic PMs of fields with two epochs of \hst\ observations by measuring the motion of stars observed by \gaia\ against the average of the field over time. Along with the method, we present tests using this application on five targets -- two satellites of the MW and three satellites of M31.

Our MW satellite targets, Sculptor and Draco, have been chosen to verify the accuracy of this method for nearby fields. Sculptor and Draco are classical dwarfs ($M_{\star} \sim 10^7 {\rm\, M_{\odot}}$) located at similar heliocentric distances (86 and 76 kpc; \citealt{mcconnachie2012}) and have been a prime focus of PM experiments, dating back to the mid-1990s (\citealt{sculptorFirstPM} for Sculptor; \citealt{dracoFirstPM} for Draco). In the last two decades, the precision of these measurements has been immensely advanced with the use of \hst\ data \citep[e.g.,][]{piatek+2006, pryor+2015, Sohn+2017, Massari+18, Massari+20}, observations from the Subaru Suprime-Cam \citep[for Draco,][]{casetiidinescu+2016}, and with \gaia\ stellar PM measurements \cite[such as from][]{MartinezGarcia+2021}.

As additional test cases of the limits of our method at M31 distances, we measure PMs for NGC 147, NGC 185, and Andromeda VII (And VII). NGC 147 and NGC 185 are relatively bright dwarf elliptical (dE) galaxies sitting comfortably within the halo of M31 (107 and 154 kpc from M31; \citealt{mcconnachie2012,savino+2022}) with ACS/WFC-based PMs and orbital histories presented in \cite{sohn+2020}. 
Our final target, the M31 satellite And VII, lies at a distance of 231 kpc from M31 \citep{mcconnachie2012,savino+2022}. While the stellar population \citep[e.g.,][]{Navabi2021} and radial velocity \citep{Kalirai2010} of this galaxy have been studied using \hst\ data, it does not have a measured PM. And VII offers an attempt to utilize archival data of overlapping fields to recover a novel result that would increase the number of M31 dwarf PMs measured using \hst\ by 50\%.

\section{Observations} \label{sec:obs}

The \hst\ ACS/WFC observations of Sculptor and Draco that we use to measure PMs are for the 3 fields of Draco (F1, F2, F3) and 2 fields of Sculptor (F1, F2) that were also used by \cite{Sohn+2017}. For Draco, the first-epoch observations were taken in 2004 October (GO-10229; PI: S. Piatek) in the F606W filter (broad V) and the second-epoch observations were taken in 2013 October (GO-12966; PI: R. van der Marel) with the same filter, pointing, and orientation as the first epoch. The observations for Sculptor were taken in 2002 September (GO-9480; PI: J. Rhodes) and 2013 September (GO-12966) in the F775W filter (SDSS i), also with the same pointing and orientation as the first. This matching of the filter, pointing, and orientation between epochs is useful especially for PM experiments, as it may help minimize the effects of instrumental- and filter-based systematics. Exposure times for all of these observations were on the order of 500 seconds. This is helpful for our purposes, as it ensures that the majority of observed \gaia\ sources are not saturated on the detector while still being photometrically deep enough for an adequate sample of stars in the chosen target.

Similarly, NGC 147 and NGC 185 have observations that were specifically set up with measuring PMs in mind; i.e., second-epoch observations with pointings and orientations identical to the first as well as identical exposure times and filters between epochs. First-epoch observations for NGC 147 were taken in 2009 November (GO-11724; PI: M. Geha) and second epoch observations in 2017 November (GO-14769; PI: S. T. Sohn). For NGC 185, the first epoch was taken 2010 January (GO-11724) and the second epoch 2017 December (GO-14769). All observations that we use were taken in the F606W filter.

These observations of NGC 147 and NGC 185 have both benefits and drawbacks when it comes to the goals of our experiment; while their setup was specifically designed for the recovery of these galaxies' PMs, the PM experiment in mind was to use background sources. Therefore, long exposure times ($t_{\rm exp} \approx 1400{\rm\, s}$) were chosen in order to maximize the S/N of these background sources in every image taken. This means that while we do have deep photometry for the galaxies in question, we also find that all of the bright, foreground MW stars are saturated on the detector. Therefore, successfully recovering a PM with these data requires being able to perform accurate astrometry on sources multiple magnitudes above the detector's saturation threshold.

The other M31 satellite in our sample, And VII, has observations that were not optimized for PM measurements. And VII has first epoch ACS observations in the F555W filter (Johnson V) taken in 2006 June (GO-10430; PI: T. Armandroff) and second-epoch observations from 2020 August (GO-15902; PI: D. Weisz) in F606W. In addition to being in slightly different band-passes, these observations differ in pointing, orientation, and exposure time. However, the fields from the two epochs do have considerable overlap.
Exposure times for the first epoch images are 600 seconds and for the second epoch images are $\sim$1100 seconds. Accordingly, we find that none of the foreground \gaia\ stars are saturated in the first epoch of images but only a handful remain unsaturated in the second set.

\section{Data Processing}

{\sc hst1pass} ({\sc h1p}; \citealt{hst1pass}) is a PSF photometry and astrometry tool that enables finding point-sources within HST {\sc flt/flc} images. {\sc flc} images, obtained from the Mikulski Archive for Space Telescopes (MAST) \hst\
search,\footnote{\href{https://mast.stsci.edu/search/ui/\#/hst}{https://mast.stsci.edu/search/ui/\#/hst}} are pre-processed to be bias- and dark-subtracted, flat-fielded, and corrected for \hst's charge transfer efficiency (CTE) effects. \hop\ finds sources, taking into account filter-dependent geometric-distortion corrections (GDCs) for the instrument used to capture the images, and fits for sources using the proper empirically-derived point-spread functions (ePSFs). \hop\ then returns both raw $(x,y)$ and distortion-corrected $(X,Y)$ detector positions of the centres of each fit ePSF, right ascensions (RA) and declinations (DEC) from the distortion-corrected World Coordinate System header for each image, along with the object's instrumental magnitude.

We then apply a rotation to the $(X,Y)$ coordinates,
\begin{equation} \label{eq:rotation}
\begin{split}
    X_{\rm new} &= \Delta X \cos{(\mathcal{O})} - \Delta Y \sin{(\mathcal{O})}, \\
    Y_{\rm new} &= \Delta X \sin{(\mathcal{O})} + \Delta Y \cos{(\mathcal{O})},
\end{split}
\end{equation}
where $\Delta X = X - a_1$ and $\Delta Y = Y - a_2$, and $a_1$, $a_2$, and $\mathcal{O}$ represent the variables {\sc POSTARG1}, {\sc POSTARG2}, and {\sc PA\_APER}, respectively, from the {\sc flc} image headers. This operation rotates the $(X,Y)$ coordinate frame around the image target so that $X$ is aligned with the sky-west direction (i.e., along RA) and $Y$ is aligned with the sky-north direction (along DEC). A benefit of performing this operation now rather than while constructing the PMs is that it forces the scatter in the $X$ and $Y$ directions (after source matching and transforming the images) to more directly correspond with the uncertainties in $\mu_{\alpha}^*$\footnote{$\mu_{\rm \alpha}^* = \mu_{\rm \alpha} \cdot \cos{(\delta)}$. This takes into account that the lines of RA converge towards the poles, and puts the units on the same scale as $\mu_{\delta}$ for all values of RA.} and $\mu_{\delta}$.

\begin{figure}
    \includegraphics[width=\columnwidth]{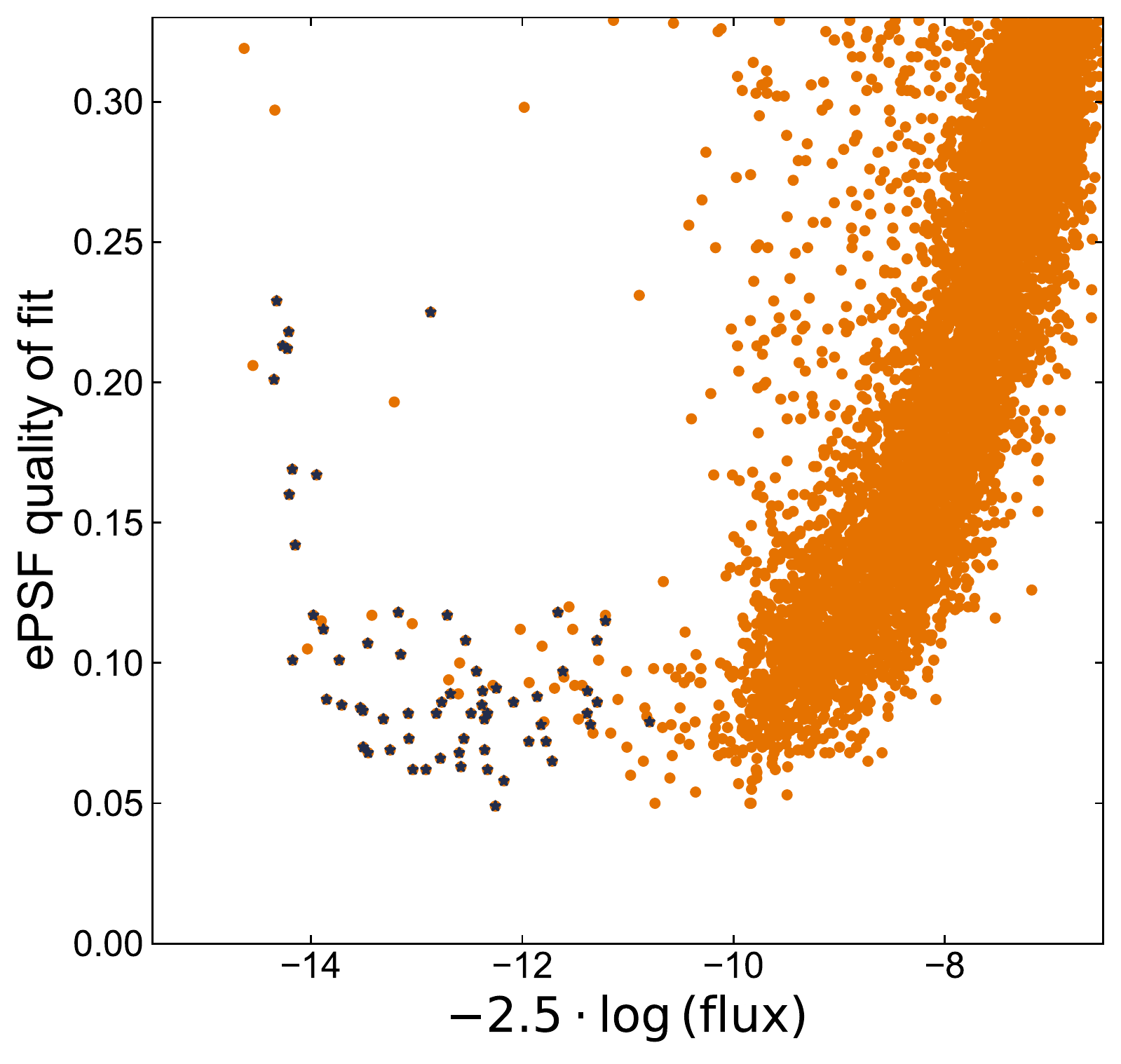}
    \caption{ePSF quality of fit as a function of instrumental magnitude from the Andromeda VII first-epoch F555W observations. Navy blue asterisks mark utilized \gaia\ foreground sources in this image.}
    \label{fig:qvspos}
\end{figure}
In order to use \hop\ to get source positions, you must choose the proper starting ePSF and GDC files. In general, it is best to choose the ePSF and GDC that match the filter and \hst\ service-mission of the image being processed. However, there are not ePSFs available for every filter; for instance, the F555W filter, which was used for the first-epoch And VII observations and for Draco F3, does not have an ePSF. In these scenarios, it is advised to use the ePSF for the closest filter (F606W for F555W, in our case). Running \hop\ with {\sc PERT5=YES} allows the ePSF to be perturbed over the image being processed on a five-by-five grid, adapting the starting ePSF to the filter being used. In general, this setting is advisable even when the ePSF matches the image's filter, as this perturbation also accounts for time-dependent and image-specific anomalies in the PSF.
In general, however, each set of images being processed requires some degree of trial and experimentation. We have also found, for instance, that using the F814W ePSF (broad I-band) for images taken in F775W results in better fits than using the F775W-specific ePSF. 

For analysing a run of \hop, the output provides a parameter that characterizes the quality of the ePSF fit, $q$, where approaching $0$ indicates a better fit, with $q \lesssim 0.2$ for confidently star-like sources \citep{hst1pass}.
$q$ is a useful diagnostic both to check the quality of astrometry returned from a run of \hop\ and for setting bounds in the subsequent steps for stars that are more likely to be foreground and for what stars should be included in image transformations.
In the case of And VII, the $q$ parameter seems to show that there is good conversion between the PSF fit and the image, with $q$ then increasing as we approach the saturation threshold just past $m_{\rm instr} \simeq -14$, which also corresponds with an increase in overall position scatter (Figure \ref{fig:qvspos}).

After constructing this source list, we run a query of the \gaia\ archive for stars with \gaia\ eDR3 data that are within the stamp on the sky covered by our images' fields-of-view. Stars selected must also fit the \gaia\ parameters:
\begin{verbatim}
    ruwe < 1.1,
    ipd_gof_harmonic_amplitude <= 0.2,
    visibility_periods_used >= 9,
    astrometric_excess_noise_sig <= 2,
    pmra_error < 1,
    pmdec_error < 1,
    parallax_error < 0.7,
\end{verbatim}
which ensures that we are choosing stars that are unlikely to be binaries and have high-quality measurements for their stellar PMs.
This list is then cross-matched on ${\rm (RA,DEC)}$ with the source lists from \hop\ after propagating the \gaia\ stellar positions for each epoch using the \gaia\ ADQL function {\sc EPOCH\_PROP\_POS}.

In addition, we search the target field against objects in the Million Quasars Catalog \citep{milliquas2021}.
Because QSOs are of relatively meager numbers compared to stars or galaxies, this search fails to return any matches for most fields; of our 6 targets, only the fields of Draco contained QSOs from this catalogue (in fact, these fields were originally targeted on QSOs). However, since these sources have no PMs of their own (i.e., are approximately stationary on the sky), fields where these sources do appear benefit greatly, as the only source of uncertainty for these objects is from the \hst\ centroiding and transformation. 

Using a field's total list of stars and their distortion-corrected $(X,Y)$ positions in each image, the original \hop\ coordinates of all stars $(X_{\rm in},Y_{\rm in})$ are transformed by fitting for the parameters in the 6-dimensional transformation:
\begin{equation} \label{eq:6dtrans}
\begin{split} 
    X_{\rm new} &= A_X + B_X \cdot X_{\rm in} + C_X \cdot Y_{\rm in} \\
    Y_{\rm new} &= A_Y + B_Y \cdot X_{\rm in} + C_Y \cdot Y_{\rm in},
\end{split}
\end{equation}
with the $A$ terms capturing translational offsets and the $B$ and $C$ terms capturing the magnification, skew, and rotation of $X$ and $Y$.
An independent fit using Equation \ref{eq:6dtrans} is made for each image using its member list,\footnote{Our script for this transformation makes use of the {\sc mpfit} function from \href{https://github.com/segasai/astrolibpy}{https://github.com/segasai/astrolibpy}, an adaptation from the {\sc Fortran} routine of the same name to {\sc Python}.} and then stars are cut from all lists for which the average standard deviation of the star's separation from the reference location in all images is higher than a set tolerance. This fit-and-cut is repeated, lowering the tolerance each time until it reaches 0.05 pixels.
Note that this means that a single image must have, at the last step, at least six members that are also in the reference image in order to do this transformation; however, those 6 members do not necessarily have to be shared with any of the other images. Requiring a source to be in at least $90\%$ of the images ensures some degree of overlap while also budgeting for a variety of issues (e.g., interference from cosmic rays) that would exclude a star from one of the catalogs.

There are multiple paths available for restricting or determining stellar membership for a target to cut down the source list before this step, each with its own advantages and shortfalls. For most well-populated targets, the fields are considerably dense enough with members that we are able to disregard membership at this step, with non-member contributions to the image transformations being $\sigma$-clipped out with each iteration of our frame transformation. With this approach, using cuts on $q$ (informed by figures such as Figure \ref{fig:qvspos}) does very well to limit the transformation to stars that are target members. For "nearby" fields, using \gaia\ stars that are likely members of the target may be a valid method to verify membership (e.g., identifying members using {\sc GetGaia}; \citealt{getgaia1}, \citealt{MartinezGarcia+2021}), but this will restrict the number of stars available for the transformation. If multiple filters are available, you may also cut membership based on color-magnitude diagram positions. However, this may require more optimized photometry than is required for the astrometric PM measurement.

The next step is to find the individual PMs for the foreground \gaia\ stars that we identified in the image. To do this, we find the median pixel position of a given foreground star in the first and second epoch and take the standard deviation of these positions as the uncertainty for each epoch. A star's total displacements in the north and east directions then become, taking into account the ACS/WFC scaling of $1 {\rm\, pixel} = 50 {\rm\,mas}$,
\begin{equation} \label{eq:totdis}
\begin{split}
	dE &= 50\frac{\rm mas}{\rm pix} \cdot \frac{(X_{\rm E1} - X_{\rm E2})}{\Delta T}, \\
	dN &= 50\frac{\rm mas}{\rm pix} \cdot \frac{(Y_{\rm E2} - Y_{\rm E1})}{\Delta T}.
\end{split}
\end{equation}
The time baseline, $\Delta T$, is the time between the start of the first image exposure in the first epoch and the first image exposure in the second epoch. These values $dE$ and $dN$, in ${\rm mas/yr}$, are the addition of the reflex motions of the target (which was transformed to be the stationary frame of reference) and the star's own PM ({\sc pmra} and {\sc pmdec} from \gaia). The PM of the target (as indicated by an individual foreground star) is therefore
\begin{equation} \label{eq:pms}
\begin{split}
	\mu_{\rm \alpha}^* &= -(dE - {\textsc{pmra}}), \\
	\mu_{\rm \delta}   &= -(dN - {\textsc{pmdec}}).
\end{split}
\end{equation}
Uncertainties are propagated through to these values, which include the uncertainty in the $(X,Y)$ transformation to the reference frame (the standard deviation of a member's coordinates in all images relative to the reference image), the uncertainty in the position of the \gaia\ star in a given epoch, and the uncertainty on the \gaia\ {\sc pmra} and {\sc pmdec}.

Finally, for targets with a large number of foreground sources, the PM of the target is reported by finding the inverse-variance weighted median of the individual $(\mu_{\rm \alpha}^*,\mu_{\rm \delta})$ results, which is found by sorting the PMs along each axis, smallest-to-largest, and finding the first PM in this list, $\mu[j]$, for which
\begin{equation}\label{eq:wmed}
    \frac{\sum\limits_{i=0}^{i=j} \sigma_i^{-2}}{\sum\limits_{i=0}^{i=n} \sigma_i^{-2}} \geq 0.5,
\end{equation}
and then taking the weighted-mean PM over this boundary as the median result.
Lower and upper 68\%-confidence uncertainties are then found using the same method, for which the left-hand side of Equation \ref{eq:wmed} is $\geq 0.15865$ and $\geq 0.84135$, respectively. For fields where we suspect a sub-sample of foreground stars to be under the influence of large systematics or that lack an adequate number of sources to calculate the weighted-median, we instead calculate the weighted average.

\section{Results} \label{sec:results}
\subsection{Sculptor and Draco} \label{ssec:mwd}
\begin{figure*}
    \includegraphics[width=\textwidth]{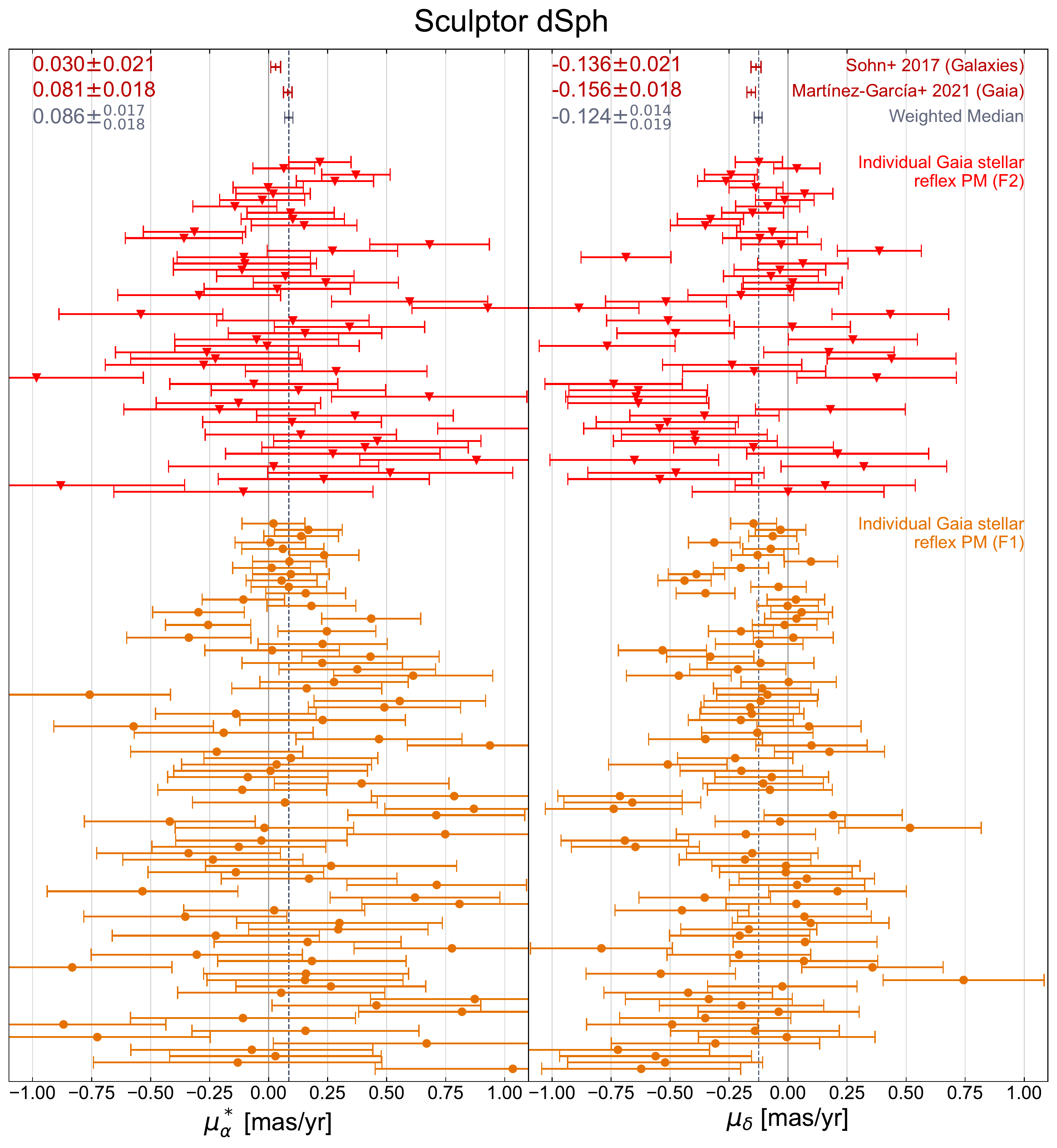}
    \caption{Summary of the results for the proper motion study of Sculptor. The left and right panels show the results for $\mu_{\rm \alpha}^*$ and $\mu_{\delta}$, respectively. The orange points (circular markers) at the bottom represent the individual reflex-PMs for each foreground \gaia\ star in the F1 field, and the red (triangular markers) points above these are the same for the F2 field. Within each field, the PMs for individual stars are sorted by \gaia\ $G$ with the brightest star starting at the top. The blue point (third from the top) is the inverse-variance weighted median of the individual PMs with $68\%$ confidence uncertainty, with the dotted line tracing the median throughout the entire plot. The top two scarlet points are the proper motion for Sculptor as reported by \protect\cite{Sohn+2017} using background galaxies in \hst\ images and by \protect\cite{MartinezGarcia+2021} using \gaia.}
    \label{fig:sculptorsummary}
\end{figure*}
\begin{figure*}
    \includegraphics[width=\textwidth]{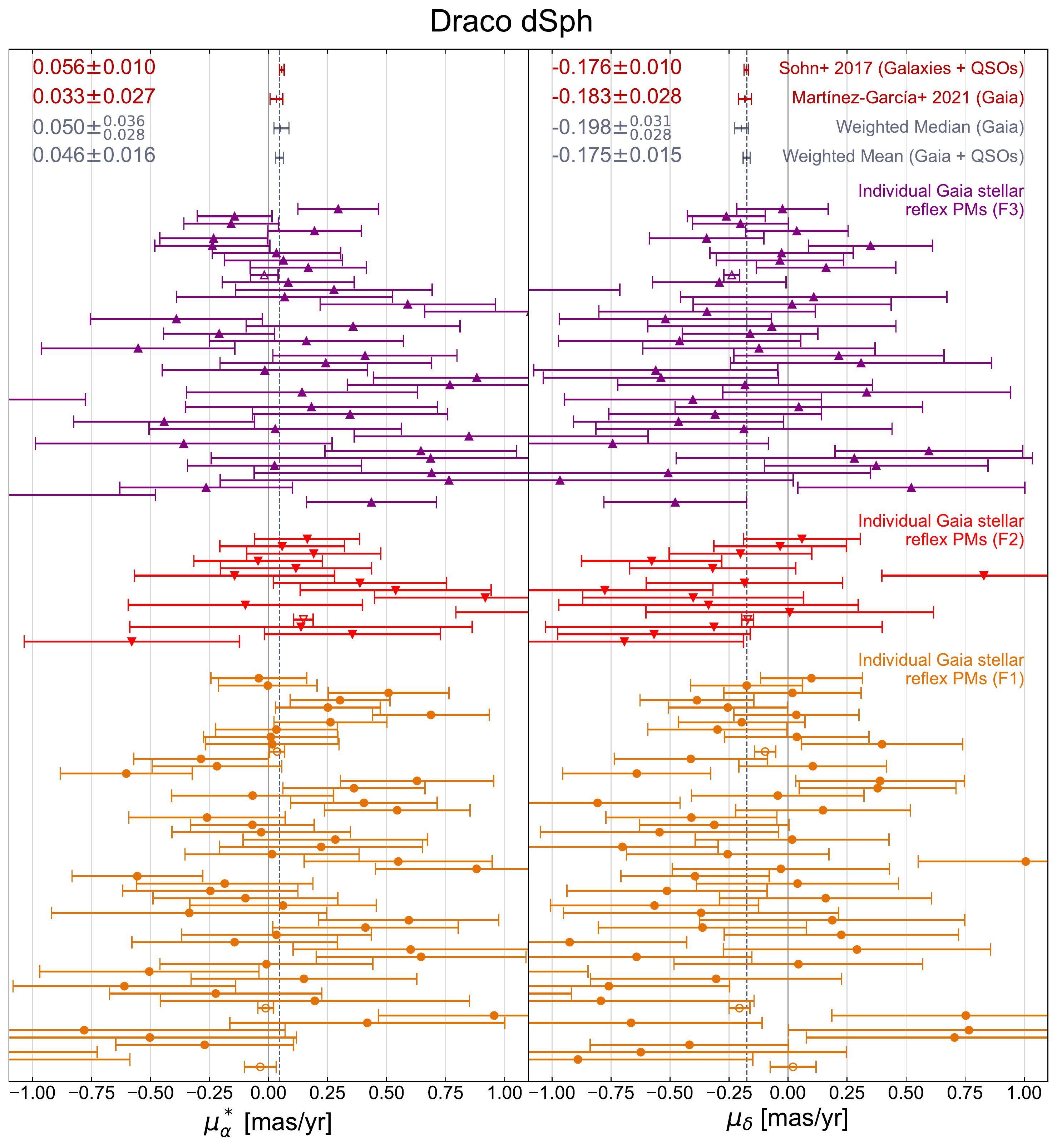}
    \caption{Summary of the results for the proper motion study of Draco. The left and right panels show the results for $\mu_{\rm \alpha}^*$ and $\mu_{\delta}$, respectively. The orange points (circular markers) at the bottom represent the individual reflex-PMs for each foreground \gaia\ star in the F1 field, the red (downwards-triangular markers) points above these are the same for the F2 field, and the purple (upwards-triangular markers) for the F3 field. Open markers represent QSOs observed in the fields. Within each field, the PMs for individual stars are sorted by magnitude with the brightest star starting at the top. The blue points (third and fourth from the top) are the inverse-variance weighted median and mean of the individual PMs with $68\%$ confidence uncertainty, excluding and including the field QSOs, respectively. The top two scarlet points are the proper motion for Sculptor as reported by \protect\cite{Sohn+2017} for and by \protect\cite{MartinezGarcia+2021}.}
    \label{fig:dracosummary}
\end{figure*}
Our results for the PMs for Sculptor and Draco are summarized in Figures \ref{fig:sculptorsummary} and \ref{fig:dracosummary}, respectively. In addition to the steps outlined in the previous section, the final foreground source lists for these objects were iterated through a 3-$\sigma$ clip on $\mu_{\alpha}^*$ and $\mu_{\delta}$. The main effect of this cut is to eliminate sources that were either mismatched between \hst\ and \gaia\ or that have low enough S/N such that they deviate significantly from the median. In these figures, we compare our per-star and weighted-median values for the PMs to some of the values found in the literature (\citealt{Sohn+2017} and \citealt{MartinezGarcia+2021}).
\cite{Sohn+2017} used only background galaxies to measure the PM of Sculptor and background galaxies and QSOs to measure the PM of Draco. On the other hand, the PMs from \cite{MartinezGarcia+2021} are solely based on the average motions of these galaxies' members found in \gaia\ eDR3.\footnote{We have chosen \cite{MartinezGarcia+2021} over \cite{Battaglia+2021} or \cite{li+2021} for their treatment of systematic uncertainties, which are the dominant source of uncertainty, especially for these classical dwarfs.}

For Sculptor, we find the values
\begin{equation}
    \begin{pmatrix} \mu_{\alpha}^* \\ \\ \mu_{\delta}^{\phantom{*}} \end{pmatrix}_{\rm Sculptor}
        =
    \begin{pmatrix} \phantom{-}0.0775 \,\pm\,^{0.0180}_{0.0190} \\ \\
        -0.1184 \,\pm\,^{0.0148}_{0.0196} \end{pmatrix} {\rm mas\,yr^{-1}}.
\end{equation}
These values agree within approximately $2.7\sigma$ ($\mu_{\alpha}^*$) and $0.6\sigma$ ($\mu_{\delta}$) with \cite{Sohn+2017} and within $0.3\sigma$ and $1.7\sigma$ with \cite{MartinezGarcia+2021}, respectively.

For Draco, we have recovered values for a PM both with and without the inclusion of QSOs.
Excluding the QSOs, we find a PM for Draco of
\begin{equation}
    \begin{pmatrix} \mu_{\alpha}^* \\ \\ \mu_{\delta}^{\phantom{*}} \end{pmatrix}_{\rm Draco}
        =
    \begin{pmatrix} \phantom{-}0.0503 \,\pm\,^{0.0358}_{0.0276} \\ \\
        -0.1979 \,\pm\,^{0.0305}_{0.0277} \end{pmatrix} {\rm mas\,yr^{-1}},
\end{equation}
which agree within approximately $0.2\sigma$ ($\mu_{\alpha}^*$) and $0.7\sigma$ ($\mu_{\delta}$) with \cite{Sohn+2017} and within $0.6\sigma$ and $0.5\sigma$ with \cite{MartinezGarcia+2021}, respectively.

\begin{table*}
    \centering
    \caption{2MASS IDs of QSOs included in our analysis, along with their approximate average {\sc flc} pixel positions.}
    \label{tab:qsolist}
    \begin{tabular}{|c|c|c|c|}
        \hline
        2MASS ID & Field & Epoch 1 $(x,y)$ & Epoch 2 $(x,y)$\\
        \hline
        J172043.09+575443.2 & F1 & $(2433.4, 1537.0)$ & $(2473.3, 1526.7)$ \\
        J172052.31+575513.2 & F1 & $(2040.1, 3073.5)$ & $(2084.5, 3063.2)$ \\
        J172148.27+575805.3 & F2 & $(2045.3, 3071.8)$ & $(2088.3, 3059.0)$ \\
        J171934.42+575849.4 & F3 & $(2040.3, 3072.8)$ & $(2082.1, 3045.7)$ \\
        J171942.15+575823.1 & F3 & $(\phantom{0}722.9, 3634.2)$ & $(\phantom{0}767.3, 3609.6)$ \\
        \hline
    \end{tabular}
\end{table*}
Within the three Draco fields, we matched a total of six QSOs (three in F1, one in F2, and two in F3). Of these sources, we consider all but one in F1, which was considerably faint, with low S/N in the \hst\ data, leaving us with five total (listed in Table \ref{tab:qsolist}). Including these sources, we recover a PM of:
\begin{equation}
    \begin{pmatrix} \mu_{\alpha}^* \\ \mu_{\delta}^{\phantom{*}} \end{pmatrix}_{\rm Draco}
        =
    \begin{pmatrix} \phantom{-}0.0458 \pm 0.0159 \\ 
        -0.1754 \pm 0.0152 \end{pmatrix} {\rm mas\,yr^{-1}}.
\end{equation}
These agree within approximately $0.6\sigma$ ($\mu_{\alpha}^*$) and $0.1\sigma$ ($\mu_{\delta}$) with \cite{Sohn+2017} and within $0.5\sigma$ and $0.3\sigma$ with \cite{MartinezGarcia+2021}, respectively.

\begin{table*}
    \centering
    \caption{Summary of the average sizes of the components that contribute to the per-foreground star random uncertainty for our target fields. $\bar{\sigma}_{X,Y}$ is the mean uncertainty in the positions of the \gaia\ stars in the ACS images after the frame transformation for the first and second epochs, divided by the time baseline. $\bar{\sigma}_{\rm Gaia PM}$ is the mean of the uncertainties on \gaia's {\sc pmra} and {\sc pmdec} values.}
    \label{tab:sculptorerror}
    \begin{tabular}{|l|c|c|c|}
        \hline
        Target & Epoch 1 $\bar{\sigma}_{X,Y}$ (mas/yr) & Epoch 2 $\bar{\sigma}_{X,Y}$ (mas/yr) & $\bar{\sigma}_{\rm Gaia PM}$ (mas/yr)\\
        \hline
        Sculptor & 0.028 & 0.028 & 0.418\\
        Draco    & 0.043 & 0.039 & 0.599\\
        NGC 147  & 0.238 & 0.152 & 0.308\\
        NGC 185  & 0.198 & 0.137 & 0.236\\
        And VII  & 0.039 & 0.062 & 0.389\\
        \hline
    \end{tabular}
\end{table*}
The most dominant source of random uncertainty in our results comes from the \gaia\ stellar PM data. Table \ref{tab:sculptorerror} gives the mean size of the various components of the per-star PM uncertainties for our target fields. As it stands, being over an order-of-magnitude larger than the next largest term, nearly all of the per-star scatter in the final PM results for Sculptor and Draco can be attributed to this component. This will remain true until the mid-2020s, when the fourth \gaia\ data release should reduce the size of these PM uncertainties by a factor of $\sim$2.4 (mostly attributed to the longer time baseline; \citealt{Lindegren21a}), bringing the magnitude of this term down to the size of the uncertainty left over from the frame transformation. We can also see that the second-epoch position scatter is significantly larger than the first epoch. Though most of this uncertainty is tied to the transformation -- the residuals left behind while transforming the frame of the second epoch into the first -- we also notice that the second-epoch \hst\ images have overall poorer PSF fits than the first epoch. This is possibly due to the evolution of the ACS/WFC PSF overtime, but could also be tied to various other technical issues, such as tracking.

\subsection{NGC 147 and NGC 185} \label{ssec:ngc}
\begin{figure*}
    \includegraphics[width=\textwidth]{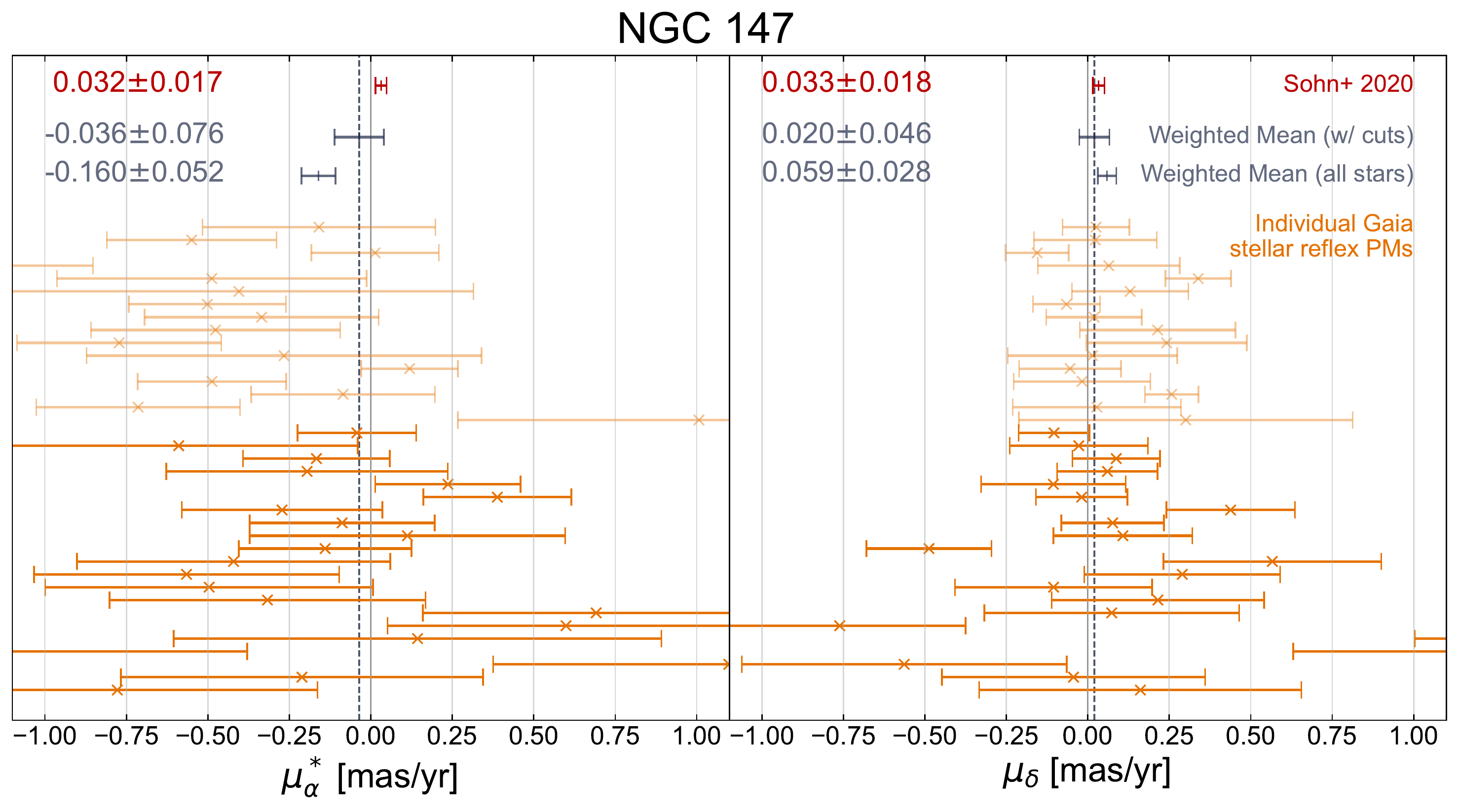}
    \caption{Summary of PM result for NGC 147. Orange points with "X" markers represent \gaia\ stars, ordered by G magnitude, that are saturated in the second epoch images, which is all of them in this case. The bottom weighted-mean uses all \gaia\ stars in the field, where the top only uses stars with instrumental magnitudes greater than -16.25 and then that pass a 3-$\sigma$ cut. Translucent orange markers are stars that failed this cut. The result from \protect\cite{sohn+2020} is provided for comparison.}
    \label{fig:ngc147}
\end{figure*}
\begin{figure*}
    \includegraphics[width=\textwidth]{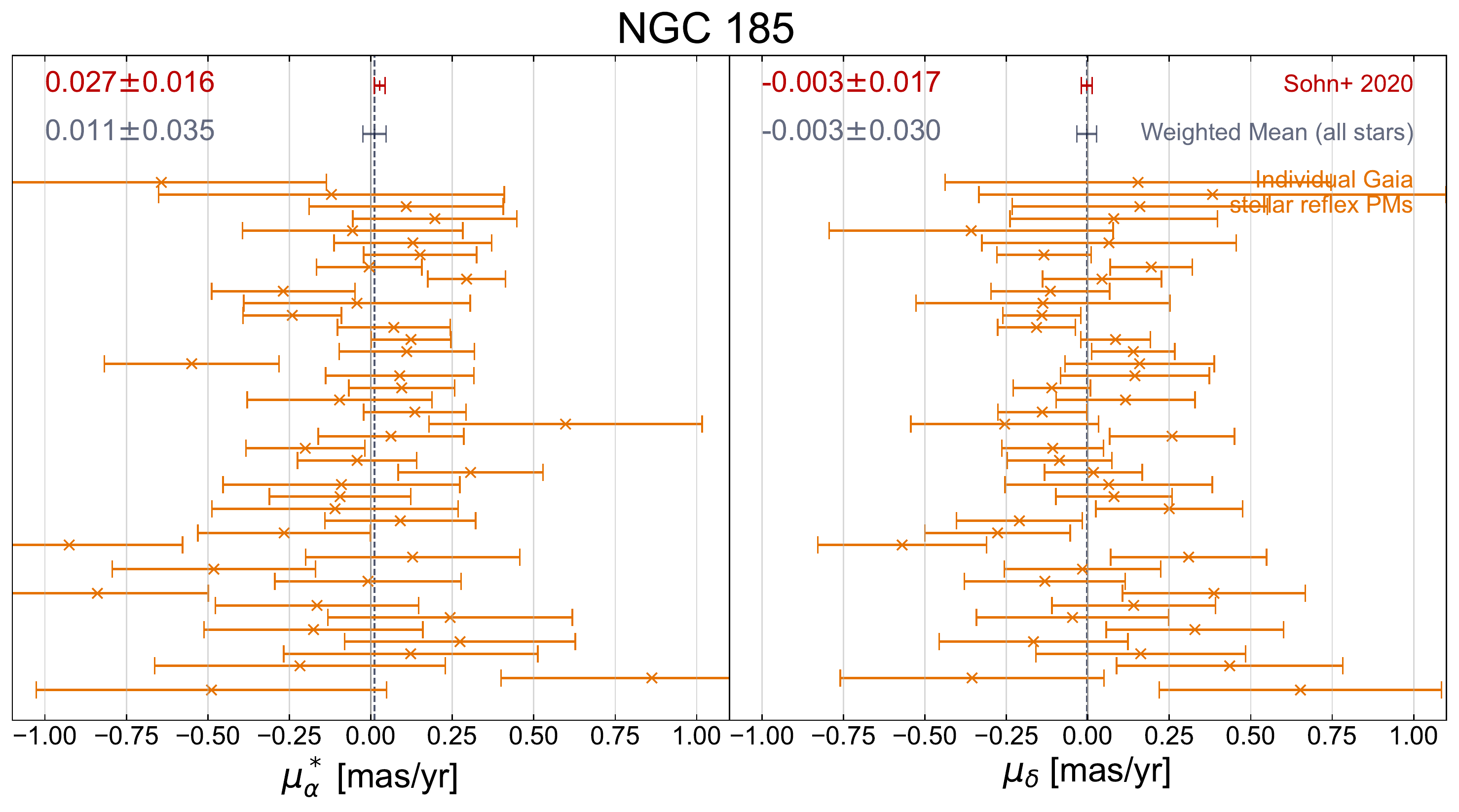}
    \caption{Summary of PM result for NGC 185. Orange points with "X" markers represent \gaia\ stars, ordered by G magnitude, that are saturated in the second epoch images, which is all of them in this case. The weighted-mean uses all \gaia\ stars in the field. The result from \protect\cite{sohn+2020} is provided for comparison.}
    \label{fig:ngc185}
\end{figure*}

Figure \ref{fig:ngc147} and \ref{fig:ngc185} summarize our results for the PMs of NGC 147 and NGC 185, respectively, versus the values reported in \cite{sohn+2020}.
The observations for both of these targets are similar in that they consist of two epochs with long exposure times ($\sim$1300 seconds), chosen both to get deep photometry of these distant galaxies as well as to get good S/N for the background galaxies used to calcualte PMs \citep{sohn+2020}. However, this also had the effect of saturating bright foreground stars on the detector; all of the stars with \gaia\ data in these fields are at least one magnitude above the WFC saturation threshold. \cite{hst1pass} shows that accurate positions may be recoverable up to six magnitudes above the saturation limit for ACS/WFC observations. However, their data does still show a trending systematic offset in the mean of the position of stars when saturated versus unsaturated, with a large scatter. Therefore, it is not obvious that the level of accuracy shown is adequate for what is required for PM experiments.

For NGC 147, the failure of the astrometry for the brightest foreground sources is obvious, with a large systematic offset from the literature value in $\mu_{\alpha}^*$. Although the scatter of the foreground reflex motions for these sources is consistent with the \gaia\ PM uncertainties, the implied PM from this result can be easily discarded as it indicates a very large, unphysical tangential velocity given the distance of NGC 147 ($\sim$600 km/s). Restricting the sample to the fainter half of stars (with instrumental magnitudes larger than $-16.0$), we are able to exclude the stars most impacted by these systematics, recovering a PM in agreement with the literature. However, this PM drags along larger random uncertainty because, being characteristically fainter than their brighter counterparts, these stars have significantly larger uncertainties associated with their \gaia\ stellar PMs.
In addition, for this smaller and more scattered sample, we must calculate the PM using an inverse-variance weighted mean, as the median is no longer operable under this domain.

For NGC 185, though done with very similar observations, we do not see the same large, obvious systematic offset that we observe for NGC 147. However, we do see an obvious increase in the uncertainties from the uncertainty on the \hst\ positions of these stars -- in fact, for some of the brightest sources, the uncertainties on stellar positions are of the same order as the uncertainties from \gaia.

\subsection{Andromeda VII} \label{ssec:and7}
\begin{figure*}
    \includegraphics[width=\textwidth]{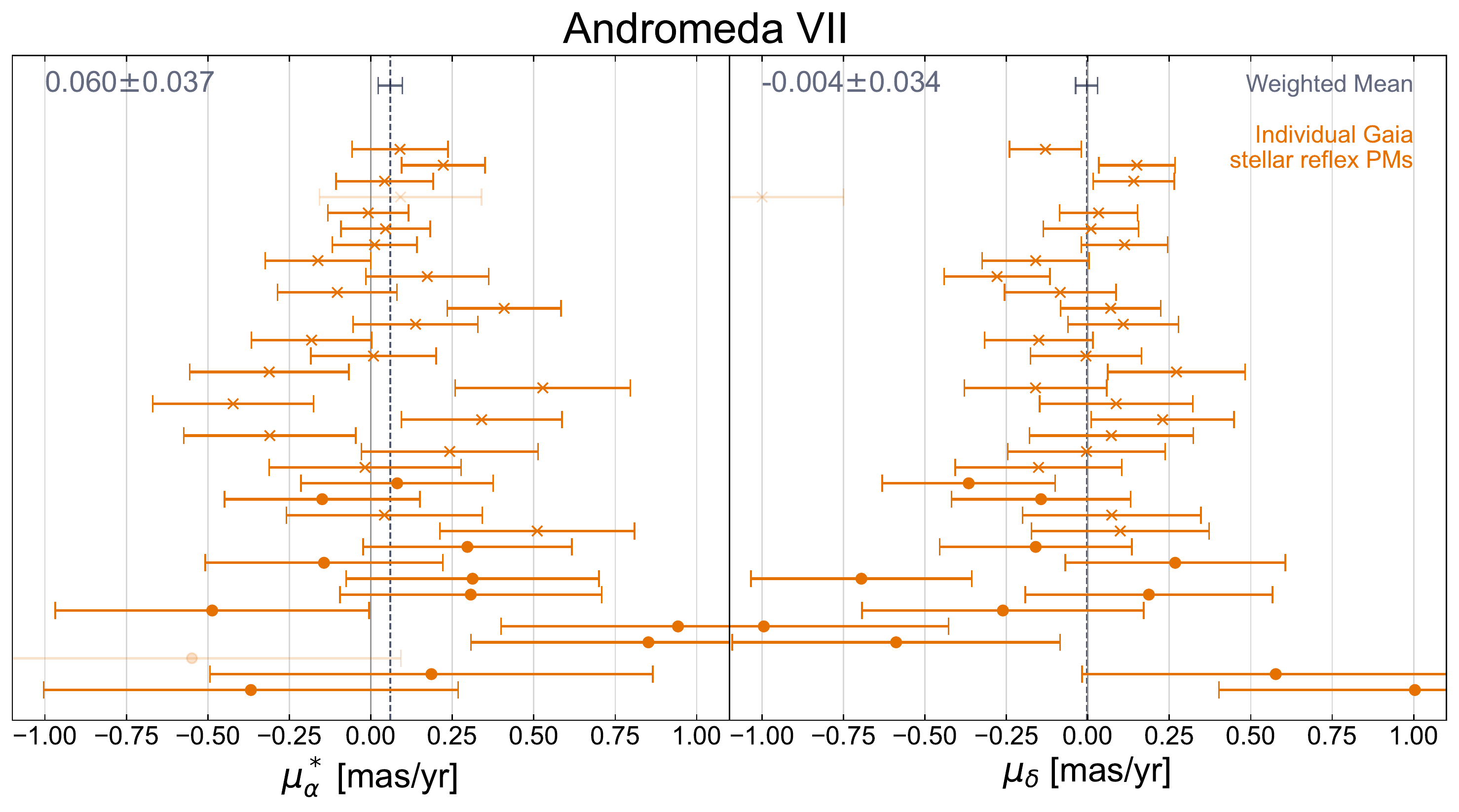}
    \caption{Summary of PM result for Andromeda VII. Orange points for individual \gaia\ stars are ordered by magnitude, with the brightest stars appearing at the top. Orange points with "X" markers represent \gaia\ stars that are saturated in the second epoch images. Orange points with circular markers are not saturated in either epoch.
    }
    \label{fig:and7summary}
\end{figure*}

Figure \ref{fig:and7summary} shows a summary of our PM result for And VII. Sundry properties of these data -- namely, the obstacles concerning the \hst\ observations of this field touched on in \S\ref{sec:obs} -- have the potential to impugn our results.
This includes that the first epoch of data is in a different filter than the second (F555W versus F606W). However, the ultimate quality of the PSF fit seems to be acceptable for these data.
Though we can have some level of confidence in our astrometric reduction for each epoch, systematic effects between the two epochs due to this filter difference (and to the differences in pointing and orientation) may remain.

The more troublesome aspect of the observations is that the second-epoch observations have very long exposure times per-image and, as a result, over half of the foreground \gaia\ stars (and, by nature, the brightest stars with the lowest \gaia\ astrometric uncertainties) are saturated. Though this poses less of an issue than it did in the cases of NGC 147 and NGC 185, this still adds uncertainty to our result, especially when considering potential biasing effects of differing orientation angles between epochs.
Though we do have a handful of foreground stars that remain unsaturated in the second epoch, restricting our calculations to only consider these stars does not improve our uncertainties as we then run into the larger \gaia\ uncertainties (as we also did when cutting the brightest stars for NGC 147). In addition, though perhaps not fully saturated, we see in Figure \ref{fig:qvspos} that the brightest stars are approaching the saturation threshold in the first epoch, possibly affecting our ability to centroid these sources.

Regardless, because the uncertainties from \gaia\ are so dominant, the increased random scatter associated with centroiding these bright stars in the \hst\ data may not have a great impact on what these stars can tell us, and it is seems likely -- especially when taking the NGC 147 and NGC 185 data into account -- that the biasing effect of saturation is realized as a systematic contribution, if at all, on the most saturated sources. Additionally, as none of the \gaia\ stars in the And VII field reach the instrumental magnitudes of the stars in the NGC 147 or NGC 185 samples, we expect this systematic to have a smaller contribution (based on \citealt{hst1pass}).
Our final PM values for And VII are therefore:
\begin{equation}
    \begin{pmatrix} \mu_{\alpha}^* \\ \mu_{\delta}^{\phantom{*}} \end{pmatrix}_{\rm And VII}
        =
    \begin{pmatrix} \phantom{-}0.0598 \pm 0.0368 \\
        -0.0037 \pm 0.0335 \end{pmatrix} {\rm mas\,yr^{-1}},
\end{equation}
which, though large, still represents a gravitationally bound and feasible speed ($\sim$235 km/s) at the location of And VII.

\subsection{Systematic errors}

Until now, we have only considered -- directly and indirectly -- random centroiding uncertainty that is caused by either photon noise or by template versus real star mismatching. There are, however, other sources of uncertainty. In particular, for \gaia, it is known that the reference frame for motion is not truly zero and that this offset varies with sky position \citep{Lindegren21a}. The offset displays spatial correlation, being more alike for similar positions as characterized by \cite{Lindegren21a} and improved for small scales in Equation 2 of \cite{Vasiliev21}. This systematic uncertainty vastly dominates the total error of many satellites of the MW, especially most of the classical ones, for which \gaia\ detects one thousand or more stars. Often, random uncertainties as small as 0.01 mas/yr or even smaller can be obtained; see, e.g., \cite{Battaglia+2021}. Using the standard deviation in the position of our reference stars to get the angular scale and calculate the systematic errors, we find that, for most targets, when only a single \hst\ field is used, the scale is about 0.014\textdegree; it increases for Sculptor to 0.1\textdegree. This is still small compared to the systematic variation scale and thus this systematic error varies only slightly between 0.0244 and 0.0261 mas/yr. Being larger than the statistical error for many targets, this term, in principle, needs to be considered as part of the error budget. Since the systematic error decreases similarly to the statistical one between DR2 and eDR3 \citep{Lindegren21a}, it is expected that the situation will be similar in the future. 

In principle, the \hst\ data could also contribute, though less is known about the systematic uncertainty of \hst. At the very least, \hst's systematics are small and in \cite{Sohn+2012, Sohn+2013} there is no evidence for systematic errors of relevant size. However, the situation could be slightly different for the QSO-based measurement of Draco in \cite{Sohn+2017} as, due to the rareness of QSOs, one cannot apply local corrections as well as one could for background galaxies.

\section{Conclusions}
\begin{table*}
    \centering
    \caption{Summary of the results for all of our targets. Distances, which are also used for calculating the PMs in physical units are from \protect\cite{mcconnachie2012} for Sculptor and Draco and from \protect\cite{savino+2022} for NGC 185 and Andromeda VII. $V_{\alpha}$ and $V_{\delta}$ values are the mean and standard deviations of draws from the asymmetric uncertainty distributions of the angular PM values.}
    \label{tab:totalsum}
    \begin{tabular}{|l|c|c|c|c|c|}
        \hline
        Target & Heliocentric Distance (kpc) & $\mu_{\alpha}^*$ (mas/yr) & $\mu_{\delta}$ (mas/yr) & $V_{\alpha}$ (km/s) & $V_{\delta}$ (km/s) \\
        \hline
        Draco & $76\,\pm\,6$ & $0.0470\,\pm\,_{0.0317}^{0.0371}$ & $-0.2157\,\pm\,_{0.0356}^{0.0314}$ & $19.8\,\pm\,11$ & $-70.5\,\pm\,12$ \\ \\
        Draco w/ QSOs & $76\,\pm\,6$ & $0.0458\,\pm\,0.0159$ & $-0.1754\,\pm\,0.0152$ & $16.5\,\pm\,5.9$ & $-63.2\,\pm\,7.4$ \\ \\
        Sculptor & $86\,\pm\,6$ & $0.0857\,\pm\,_{0.0180}^{0.0168}$ & $-0.1240\,\pm\,_{0.0193}^{0.0141}$ & $34.6\,\pm\,7.5$ & $-52.3\,\pm\,7.8$ \\ \\ \\
        NGC 185 & $648.6\,\pm\,18$ & $0.0113\,\pm\,0.0354$ & $-0.0033\,\pm\,0.0299$ & $34.9\,\pm\,110$ & $-10.3\,\pm\,92$ \\ \\
        Andromeda VII & $824.1\,\pm\,23$ & $0.0598\,\pm\,0.0368$ & $-0.0037\,\pm\,0.0335$ & $234\,\pm\,140$ & $-14.2\,\pm\,130$ \\ \\
        \hline
    \end{tabular}
\end{table*}

Our results in \S\ref{sec:results} (also summarized in Table \ref{tab:totalsum}) suggest that using foreground sources to measure the PMs of distant objects has much promise. In particular, it is encouraging that we are able to recover PMs for Sculptor and Draco that are in agreement with the literature, along with a new, physically-feasible motion for And VII. However, our results for the M31 targets do force us a consider a few important caveats:
\begin{enumerate}
    \item The most dominant source of uncertainty is from the \gaia\ stellar PMs. Therefore, this method will work best for fields where there are many foreground stars with ${\textsc{pm\_error}} \lesssim 0.5$, which should correspond to foreground stars with $G \lesssim 20$. Regardless, plans for future PM experiments should consider that the size of these uncertainties should drop by a factor of about 2.4 \citep[e.g.,][]{Lindegren21a} circa 2025 with \gaia\ DR4.
    \item The main obstacle to using this method is the quality of the \hst\ astrometric reduction that is possible given the foreground stars in a field. For more distant targets (and perhaps fainter, nearby targets such as UFDs), archival data tends to have longer exposure times, meaning that foreground stars are more likely to be saturated on the detector. For instance, though the the MW Cosmology Treasury Program (\hst\ GO-14734, PI: N. Kallivayalil; \citealt{MWtreasury}) observed 32 targets around the MW, these archival observations all have per-image exposure times of $\sim$1200 seconds, chosen in order to obtain deep photometry (e.g., as done by \citealt{Sacchi2021} and \citealt{richstein+2022} with this data). \cite{hst1pass} shows -- and we were able to confirm -- that, for stars approaching and exceeding the saturation threshold (at $m = -2.5 \log{({\rm flux})} \simeq -14$), the quality of the astrometry that is possible quickly degrades. Though \cite{hst1pass} suggests that accurate astrometry is possible up to around 6 magnitudes above the saturation threshold, we do not find this to be true for the precision required for PM studies. 

    Our suggestion is to aim for the \gaia\ stars in a target field to have instrumental magnitudes no more than $\sim$1.25 magnitudes above the saturation threshold, i.e., $m \gtrsim -15.4$; this approximately corresponds to the instrumental magntiude of the brightest stars that we included in our analysis of And VII. When planning new observations, then, the exposure time at which half of the \gaia\ stars will be over this limit (in the F606W filter) can roughly be estimated as:
    \begin{equation}
        t_{\rm exp} \approx 1000{\rm\, s} \cdot 10^{0.4 (\widetilde{G}-18.3)},
    \end{equation}
    where $\widetilde{G}$ is the median $G$-band magnitude of \gaia\ stars in the target field.
\end{enumerate}

Along with its application with existing \hst\ data, the future of space-based observatories looks promising to keep this method as an efficacious tool moving forward. The method we have presented in this paper for recovering novel PMs is well positioned to be considered for new second-epoch follow-ups to existing \hst\ observations, as it does not ask for deep-photometry/long-exposure first-epoch observations (rather, it may flourish with the obverse).

In fact -- with proper foresight -- the future should be promising for PM experiments generally. Even assuming a pessimistic outlook for \hst's lifespan, the {\it James Webb Space Telescope} and the {\it Nancy Grace Roman Space Telescope} ({\it NGRST}) should ensure our ability to execute second-epoch observations well into the 2030s. It is also exciting to consider -- with its planned 2027 launch -- that the combination of {\it NGRST}'s large field-of-view (${\sim}100$ times larger than \hst's ACS/WFC, proportionally allowing for ${\sim}100$ times more foreground stars per target field) and the improved uncertainties from future \gaia\ data releases has the potential to substantially improve the precision with which we can measure systemic PMs.

~\\
\section*{Acknowledgements}
This paper is dedicated to Dougie Gross, who spent the summer of 2021 on sabbatical at UVA. 

We would like to thank Joel C. Zinn for many valuable conversations. 

Support for this work was provided by NASA through grant GO-15902.

A. del Pino acknowledges the financial support from the European Union-NextGenerationEU and the Spanish Ministry of Science and Innovation through the Recovery and Recovery and Resilience Facility project J-CAVA.

This work has made use of data from the European Space Agency (ESA) mission
{\it Gaia} (\url{https://www.cosmos.esa.int/gaia}), processed by the {\it Gaia}
Data Processing and Analysis Consortium (DPAC,
\url{https://www.cosmos.esa.int/web/gaia/dpac/consortium}). Funding for the DPAC
has been provided by national institutions, in particular the institutions
participating in the {\it Gaia} Multilateral Agreement.

\section*{Data Availability}
The {\sc HubPUG} software can be found at and downloaded from \href{https://github.com/jackwarfield/HubPUG.git}{https://github.com/jackwarfield/HubPUG.git}. This repository also includes examples for recreating the results from this paper.



\bibliographystyle{mnras}
\bibliography{bibliography.bib} 


\bsp	
\label{lastpage}
\end{document}